\documentclass[letter,longauth]{aa}
\usepackage{natbib}
\bibpunct{(}{)}{;}{a}{}{,} 
\usepackage{graphicx}
\usepackage{hyperref}
\usepackage{xcolor}
\usepackage[varg]{txfonts}

\usepackage{amsmath}	
\usepackage{orcidlink}
\usepackage{longtable}
\usepackage{placeins}

\newcommand{\lsolar}{L$_{\odot}$}

\begin{document} 

\title{JWST detects a dusty AGB-like source before the Type Ia-CSM supernova 2026sqf}

\author{
\orcidlink{0000-0003-4610-1117}T. Szalai\inst{1,2} \and
\orcidlink{0000-0002-0763-3885}D. Milisavljevic\inst{3,4} \and
\orcidlink{0009-0006-5127-8290}N. Zimmer\inst{3} \and
B. Garretson\inst{3} \and
\orcidlink{0000-0001-8385-3727} T. Moore\inst{5} \and
\orcidlink{0000-0001-9038-9950}S.~D.~Van Dyk\inst{6} \and
\orcidlink{0000-0003-2877-5935}A. Lu\inst{7,8} \and
\orcidlink{0000-0003-2132-5632}S. Topal\inst{9} \and
\orcidlink{0000-0003-2238-1572}O.~D. Fox\inst{5} \and
\orcidlink{0000-0002-5477-0217}T. Kangas\inst{10,11} \and
\orcidlink{0000-0001-7497-2994}S. Mattila\inst{11,12} \and
\orcidlink{0000-0003-4254-2724}A. Reguitti\inst{13,14} \and
\orcidlink{0009-0002-7560-1903}U. Pylypenko\inst{10,11} \and
C. Cynamon\inst{15} \and
\orcidlink{0000-0002-1066-6098}T.-W. Chen\inst{16} \and
\orcidlink{0000-0002-9928-0369}A. Aryan\inst{16} \and
D. Caudill\inst{3} \and
D. Dickinson\inst{3} \and
\orcidlink{0000-0003-4980-1012}M. Bureau\inst{17} \and
\orcidlink{0000-0001-5033-7208}W. Choi\inst{18} \and
\orcidlink{0000-0003-4932-9379}T.~A. Davis\inst{19} \and
\orcidlink{0000-0001-6803-2138}D. Haggard\inst{8} \and
T.~M. Reynolds\inst{11,20,21} \and
\orcidlink{0000-0002-5571-1833}M. Stritzinger\inst{22} \and
P. Wiggins\inst{23}
}
     
\institute{
Department of Experimental Physics, Institute of Physics, University of Szeged, D{\'o}m t{\'e}r 9, 6720 Szeged, Hungary \\
\email{szaszi@titan.physx.u-szeged.hu}
\and
MTA-ELTE Lend\"ulet "Momentum" Milky Way Research Group, Szent Imre H. st. 112, 9700 Szombathely, Hungary
\and
Department of Physics and Astronomy, Purdue University, 525 Northwestern Ave, West Lafayette, IN 47907, USA
\and
Institute for Physical Artificial Intelligence, Purdue University, West Lafayette, IN 47907, USA
\and
Space Telescope Science Institute, 3700 San Martin Drive, Baltimore, MD 21218, USA
\and
Caltech/IPAC, Mailcode 100-22, Pasadena, CA 91125, USA
\and
University of British Columbia, 6224 Agricultural Rd 325 Vancouver, BC, V6T 1Z1, Canada
\and
Department of Physics, Trottier Space Institute, McGill University, 3600 University Street, Montreal, QC H3A 2T8, Canada
\and
Department of Physics, Van Yüzüncü Yıl University, Van 65080, Türkiye
\and
Finnish Centre for Astronomy with ESO (FINCA), FI-20014 University of Turku, Finland
\and
Department of Physics and Astronomy, University of Turku, FI-20014, Finland
\and
School of Sciences, European University Cyprus, Diogenes Street, Engomi, 1516 Nicosia, Cyprus
\and
INAF—Osservatorio Astronomico di Padova, Vicolo dell’Osservatorio 5, I-35122 Padova, Italy
\and
INAF—Osservatorio Astronomico di Brera, Via E. Bianchi 46, 23807, Merate (LC), Italy
\and
Supra Solem Observatory, Santa Lucia Mountains, CA, USA
\and
Graduate Institute of Astronomy, National Central University, 300 Jhongda Road, 32001 Jhongli, Taiwan
\and
Sub-department of Astrophysics, Department of Physics, University of Oxford, Denys Wilkinson Building, Keble Road, Oxford OX1 3RH, UK
\and
Department of Physics and Astronomy, McMaster University, Hamilton, ON L8S 4M1, Canada
\and
Cardiff Hub for Astrophysics Research \& Technology, School of Physics \& Astronomy, Cardiff University, Queens Buildings, The Parade, Cardiff CF24 3AA, UK
\and
Cosmic Dawn Center (DAWN)
\and
Niels Bohr Institute, University of Copenhagen, Jagtvej 128, 2200 København N, Denmark
\and
Department of Physics and Astronomy, Aarhus University, Ny Munkegade 120, DK-8000 Aarhus C, Denmark
\and
Department of Physics \& Astronomy, University of Utah, Salt Lake City, UT 84112-0090, USA
}

\date{Accepted XXX. Received YYY; in original form ZZZ}

 \abstract
   {Supernova (SN) 2026sqf recently appeared in the nearby face-on spiral galaxy NGC~3310 and has shown signs of strong interaction with a circumstellar medium (CSM). Such intense interaction, rare among nearby SNe, offers a valuable opportunity to reveal details on the origin and nature of its progenitor system.}
   {We present early-time photometric and spectroscopic observations of SN 2026sqf, along with our identification and characterization of a candidate pre-explosion counterpart in space telescope imaging.}
   {We analyzed the early-phase spectra and light curves (LCs) of SN 2026sqf. We identified a point source near the SN position in pre-explosion JWST and HST imaging, tested its association with the SN through astrometric alignment of pre- and post-explosion images and a chance-projection analysis, and constructed and modeled its spectral energy distribution (SED).}
   {The general characteristics of the spectra, the strengths of the emission lines, and the LC evolution all suggest that SN 2026sqf belongs to the rare SN Ia-CSM subclass, which would make it the closest known member (D $\sim$ 19 Mpc). The point source is a credible counterpart of the progenitor system, the first such identification for this SN subclass. Under our modeling assumptions, carbonaceous dust around a luminous evolved star provides a reasonable description of its combined SED; compared with local analogues, the candidate most closely resembles an extreme asymptotic giant branch (AGB) star.}
 {We report the first candidate pre-explosion counterpart of a thermonuclear SN identified in JWST imaging. Our results support the scenario in which SNe Ia-CSM emerge from a system consisting of a white dwarf and an evolved giant (probably AGB) star. Strong binary interaction, possibly involving common-envelope evolution or a merger, is a natural explanation for the observed properties of the SN and its candidate counterpart, though the channel and timescale remain uncertain. Late-time follow-up, especially with JWST, will test the identification and the conclusions of our early-phase analysis.}
   \maketitle
 \nolinenumbers
 
\section{Introduction}

The progenitor systems of Type Ia supernovae (SNe~Ia) remain inadequately understood \citep[see e.g.][]{MMN2014,Ruiter_2025}. Although it is well established that the explosion arises from the thermonuclear disruption of a carbon-oxygen white dwarf (WD), whether the companion is a non-degenerate donor star (the single-degenerate channel) or a second WD (double-degenerate) is still debated. 
Direct imaging of a SN~Ia progenitor prior to explosion, when achievable, provides the most powerful test of the donor's nature, but the rarity of nearby, well-imaged SN Ia sites (or, the lack of luminous companions, e.g. \citealp{Li_2011,Kelly_2014}) has to date prevented a solid identification (except for a He-rich donor star found in the special Type Iax SN~2012Z, \citealp{McCully_2014}).

SN~2026sqf presented a rare opportunity to close this gap. It was discovered on 2026-07-08.25 UT (61229.25 MJD) by P. Wiggins \citep{Wiggins_2026sqf_disc}, with a last non-detection on 2026-07-05.23 UT. It is the fourth supernova discovered in the nearby ($D \sim$ 19 Mpc) face-on spiral galaxy NGC~3310, after SNe 1974C, 1991N, and 2021gmj. This host galaxy benefits from deep, multi-epoch archival imaging extending years before the explosion, enabling direct identification of the potential progenitor system.

The first classification spectra obtained on the day of discovery show prominent narrow H and He emission lines, suggesting a Type II SN with features from flash-ionised circumstellar material (CSM; \citealp{Balcon_2026sqf_class,Wise_2026sqf_class}). 
However, based on its later photometric and spectral evolution, it has been suggested that SN~2026sqf may rather belong to the special type of SNe Ia-CSM \citep{Midavaine_2026,Rehemtulla_2026}.

SNe Ia-CSM are thought to arise from thermonuclear explosions of WDs surrounded by dense, H-rich shells of ambient matter. Their early-time spectral evolution can resemble that of normal SNe~Ia, however, later (or even sometimes in the earliest spectra) they start to show narrow emission lines as signposts of the emerging CSM interaction. 
The first famous examples of this subclass were SNe 2002ic \citep[e.g.,][]{Hamuy_2003,Livio_2003,Wang_2004,Deng_2004,Kotak_2004,Wood-Vasey_2004}, 2005gj \citep{Aldering_2006,Trundle_2008}, and PTF11kx \citep{Dilday_2012}.
The first systematic study was published by \cite{Silverman_13_IaCSM}, who also suggested the nomenclature of SNe Ia-CSM for these events. 
To date, only a few dozens of SNe Ia-CSM are known, and most of them are well beyond 100 Mpc.
Recent systematic studies of the Zwicky Transient Facility (ZTF) dataset \citep{Sharma_2023,Terwel_2025a,Terwel_2025b} verified that such events are indeed rare ($<<$1\% of the total SN Ia rate). 

Due to their rare occurrence and large distances, the origin and explosion mechanism of SNe Ia-CSM are still mostly unclear. Some studies suggest symbiotic nova-like systems -- i.e., a WD accompanied by a red giant branch (RGB) or asymptotic giant branch (AGB) star -- as progenitors \citep{Hamuy_2003,Dilday_2012,Moore_Bildsten_2012}, while others argue for WD + main sequence (MS) star pairs, which may also form a massive common envelope before explosion of the WD companion \citep[see e.g.][]{Hachisu_2008,Meng_Podsliadlowski_2017}. In some cases, even a core-collapse origin has been assumed \citep{Benetti_2006,Inserra_2014,Inserra_2016}; however, recent studies rather favor the thermonuclear origin from a binary system \citep[e.g.][]{Sharma_2023,Uno_2023b}. 

To date, no direct progenitor identification has been achieved for any SN~Ia-CSM. Except for the prior closest known (D $\sim$ 80 Mpc) SN Ia-CSM 2012ca \citep{Bochenek_2018}, and the atypical SN~2020eyj with a He-rich CSM \citep{Kool_2023}, post-explosion radio/X-ray observations provide only upper limits \citep[see][]{Dwarkadas_2025_Uni,Griffith_2025}, imposing strong constraints on our ability to reveal the pre-explosion mass-loss histories of such events. By contrast, SNe~Ia-CSM are extremely bright in the mid-infrared (mid-IR) even years after explosion, supporting the idea of dusty environments and/or indicating post-explosion dust-formation processes \citep{Fox_Filippenko_2013,Szalai_2021,Mo_2025}.

SN~2026sqf is a similarly interacting event, but at a considerably closer distance than any SN~Ia-CSM studied to date, making it an excellent target for an extensive follow-up study. Here, we present the results of early-time photometric and spectroscopic observations of SN~2026sqf, along with our efforts to identify and characterize its potential progenitor system in pre-explosion space telescope imaging.

\section{Data and analysis}

The host galaxy NGC~3310 has a redshift of $z \sim$ 0.0033 based on the NASA/IPAC Extragalactic Database (NED)\footnote{\href{https://ned.ipac.caltech.edu}{https://ned.ipac.caltech.edu}}. 
During this work, we use $D = 19.0 \pm$ 1.5 Mpc for the distance and $E(B-V)_\textrm{tot}=0.04$ mag for the total reddening (see details in Sec. \ref{sec:A_early_dist}).

\subsection{Early-time spectroscopy and photometry of SN~2026sqf}\label{sec:data_spec}

We obtained spectra on SN~2026sqf on four nights (on 07-13, 07-17, 07-22, and 07-28, Fig. \ref{fig:spec}) with the Alhambra Faint Object Spectrograph and Camera (ALFOSC) on the 2.56-m Nordic Optical Telescope (NOT, located at the Roque de los Muchachos Observatory, La Palma), as part of the NOT Un-biased Transient Survey 2 (NUTS2) program\footnote{\href{https://nuts.sn.ie/}{https://nuts.sn.ie/}}.
Details are found in Sec. \ref{sec:A_early_spec}.

We corrected our observed spectra for redshift and extinction and compared them to a large sample of SN templates of {\tt SNID-SAGE} (SuperNova IDentification-Spectral Analysis and Guided Exploration, \citealp{SNID-SAGE_2026}).
The spectra match the Ia-CSM templates best at all four epochs, and the distinction from SNe IIn grows with time:
in addition to strong H emission lines, several characteristics of overluminous SNe Ia  -- \ion{Fe}{iii}, \ion{Si}{ii}, \ion{Ca}{ii}, and weak \ion{S}{ii} -- become increasingly prominent (see also Sec. \ref{sec:A_early_spec}). 

Furthermore, as discussed in detail in \cite{Silverman_13_IaCSM} and in \cite[][hereafter S23]{Sharma_2023}, both the H$\beta$ equivalent width (EW) and the H$\alpha$/H$\beta$ intensity ratio (Balmer decrement) are good tests to separate SNe Ia-CSM from SNe IIn: in SNe Ia-CSM, H$\beta$ lines are weaker, and thus the ratio is larger). 
From the earliest up to the latest spectrum, we found EW$_\textrm{H$\alpha$} = 42.5 \pm 1.5$ and $26.2 \pm 0.3$ \AA\ and  EW$_\textrm{H$\beta$} = 8.4 \pm 0.8$ and $5.3 \pm 0.1$ \AA, which leads to a value of F$_\textrm{H$\alpha$}$/F$_\textrm{H$\beta$} \sim 5$. Both of these EW$_\textrm{H$\beta$}$ and F$_\textrm{H$\alpha$}$/F$_\textrm{H$\beta$}$ values match the median values of the SN Ia-CSM sample in S23 and differ well from those of SNe IIn (EW$_\textrm{H$\beta$}$ = 13 \AA, and F$_\textrm{H$\alpha$}$/F$_\textrm{H$\beta$}$ = 3).

We obtained early-time optical photometry of SN 2026sqf using the 0.5-m telescope at the Supra Solem Observatory (SSO), the 0.4-m Seisi Lulin Telescope (SLT) and the Lulin One-meter Telescope (LOT), as well as a 0.35-m Celestron C-14 (IAU code 718) associated with the discovery report \citep{Wiggins_2026sqf_disc}. Publicly available ZTF photometry was also collected. The SN was followed up to August 6, near maximum light, shortly before entering solar conjunction.

We compared the early-time LCs of SN~2026sqf to those of known SNe Ia-CSM from the ZTF sample, and also to a large sample of interacting Type II SNe. As shown in Fig. \ref{fig:LC_comp_ztf}, SN 2026sqf follows an LC evolution similar to that of known SNe Ia-CSM and is clearly more luminous than Type II SNe with early interaction.

\subsection{Analysis of the potential progenitor system}\label{sec:data_jwst_hst}

\begin{figure}
    \centering
    \includegraphics[width=0.7\columnwidth]{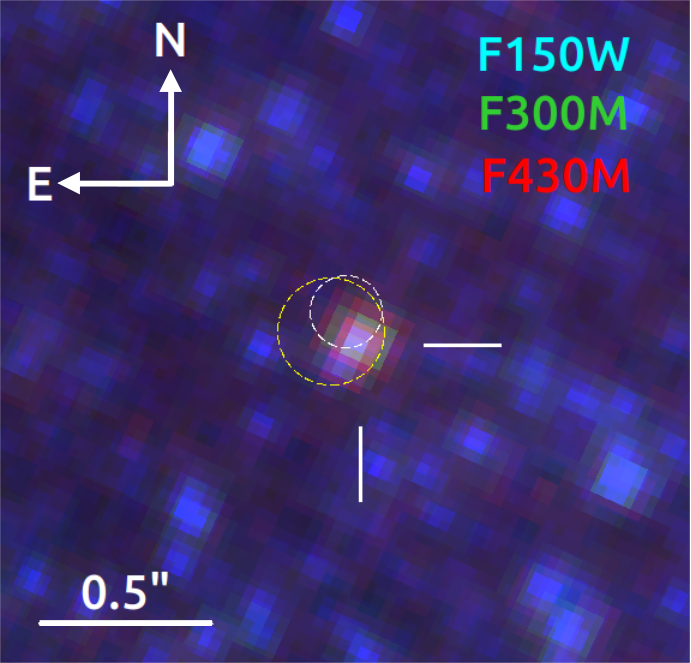}
     \caption{Color composite picture created from {\it JWST} GO 9246 NIRCam F150W, F300M, and F430M pre-explosion images (obtained in March 2026) of the site of SN~2026sqf. Yellow and white dashed circles show the uncertainties of HST WFC3 and ACS astrometric solutions, respectively.}
    \label{fig:nircam}
\end{figure}

We used the Barbara A. Mikulski Archive for Space Telescopes (MAST) to check the availability of pre-explosion {\it JWST} and {\it HST} images of the field.
There are two sets of {\it JWST} images that cover the site of SN~2026sqf. During the NIRCam GO 9246 program (PI: A. Lu), the galaxy NGC~3310 was observed with several filters in March 2026. In 2024, MIRI also observed this region during program GO 3295 (PI: D. Sand) with all nine filters between F560W and F2550W.

Data from several {\it HST} programs also cover this region; we collected and analyzed all of these images. 
We also checked the {\it Spitzer}/IRAC \footnote{\href{https://irsa.ipac.caltech.edu/applications/Spitzer/SHA/}{https://irsa.ipac.caltech.edu/applications/Spitzer/SHA/}} images of the field but found that both their spatial resolution and sensitivity are too low to resolve any point sources in the vicinity of the SN position.

After alignment and relative astrometry (Sec. \ref{sec:A_jwst_hst}), we identified a red point source on both the {\it JWST}/NIRCam and MIRI images 
at [($\alpha$(2000) = $10^{\rm hr}38^{\rm m}47.953^{\rm s}$, $\delta$(2000) = $+53^\circ30'34.08''$]. The object is clearly seen in all wide- and medium-band NIRCam filters, just as in MIRI filters up to 10 $\mu$m (Figs. \ref{fig:nircam}, \ref{fig:miri}). 

\subsubsection{JWST and HST photometry}\label{sec:data_jwst_hst_phot}

Following the strategy described in \cite{SVD_2026_25pht}, we performed photometry on {\it JWST} and {\it HST} data applying both \texttt{space{\textunderscore}phot}\footnote{https://zenodo.org/records/12100100} \citep{Pierel2024} and {\tt Dolphot} \citep{Dolphin2016,Weisz2024}. Details are found in the Appendix (Sec. \ref{sec:A_jwst_hst}).

For {\it JWST} data, we found that the values of the two independent methods agree to within $\sim$0.15 mag ($\lesssim 2\sigma$ of the uncertainties) in each case, except for the MIRI F770W filter (with a $\sim 2 \times$ larger uncertainty; see Fig. \ref{fig:spacephot}). Thus, we used the average flux values of the two methods in every filter as our final photometry (see Table \ref{tab:JWST_phot}).
Note that in the F1130W filter and at longer wavelengths, only upper limits can be established, due to the lack of a clear point source detection at the SN position.

For {\it HST} data, we applied {\tt Dolphot} for the photometry of WFPC2 images, while \texttt{space{\textunderscore}phot} was used for WFC3 and ACS data. We have a slight ($\sim$ 4 $\sigma$) detection of a source found at the same position as on {\it JWST}/NIRCam and MIRI images on the {\it HST} F814W images obtained with WFC3 and ACS. 
For all the other filters, we only have upper limits (see Table \ref{tab:HST_phot}).

\subsubsection{SED of the potential progenitor system}\label{sec:data_sed}

\begin{figure}
    \centering
    \includegraphics[width=\columnwidth]{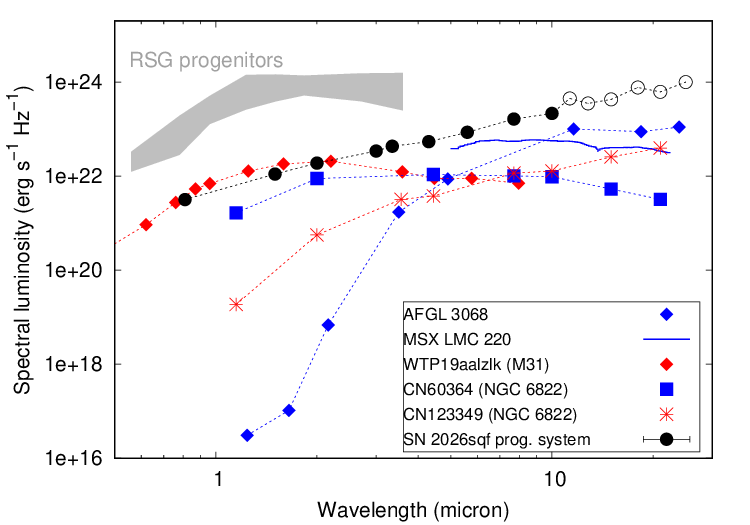}
     \caption{Comparison of the SED of the potential progenitor system of SN~2026sqf (filled and empty circles denote positive detections and upper limits, respectively) to those of the most luminous local AGB stars -- AFGL 3068 \citep{Unnikrishnan_2025}, MSX LMC 220 \citep{Sloan_2026}, CN60364 and CN123349 \citep[NGC 6822,][]{Nally_2026} --, and of the potential CEE event WTP19aalzlk in M31 \citep{Karambelkar_2025}. Blue and red symbols denote C-rich and O-rich objects, respectively. Gray region shows the luminosity region of several resolved RSG progenitors of Type II SNe \citep{VanDyk2025}.
    }
    \label{fig:sed_analogs}
\end{figure}

We constructed the SED of the potential progenitor system of SN~2026sqf from the dereddened {\it JWST} and {\it HST} fluxes (Fig. \ref{fig:sed_analogs}).

We tried to reveal the nature of the observed object/system by looking for local analogs.
By integrating the observed SED, we get luminosities of $L_\textrm{IR,NIRCam} \sim 6000 $ \lsolar\ and $L_\textrm{IR,full} \sim 17,800 $ \lsolar\ (when using only {\it JWST}/NIRCam fluxes or the full observed SED from HST F814W up to MIRI F1000W filter, respectively). This places the object well above RGB stars ($L\lesssim2500$ \lsolar; see \citealp{Sarkar_2022}) and into the regime of luminous AGB stars or, potentially, red supergiants (RSGs, \citealp{Nally_2026}); the latter, however, typically extend to substantially higher luminosities ($L\sim10^4$--$10^{5}$\,\lsolar; \citealp{VanDyk2025}), with only the lowest-luminosity RSGs overlapping our fitted value.

In Fig. \ref{fig:sed_analogs}, we compare the SED of the progenitor system candidate of SN~2026sqf to those of some of the brightest AGB stars. 
The SED of our JWST source is similar to those of the selected local analogs, but is generally even more luminous (especially in the mid-IR range). Nevertheless, it is less luminous than known RSG progenitor objects identified for nearby H-rich Type II SNe \citep{VanDyk2025}.

All of these known systems are generally inside dusty regions due to either stellar winds or episodic mass-loss events. The circumstellar environments of luminous AGB stars are typically carbon-rich \citep[e.g.,][]{Boyer_2011, Unnikrishnan_2025, Sloan_2026}, reflecting the surface carbon enrichment produced by third dredge-up during the AGB thermal-pulse phase. RSGs do not undergo this process and instead are sites of oxygen-rich, silicate dust \citep[e.g.,][]{SVD_2026_25pht}. 

To test our initial assumptions and determine the physical parameters of both the stellar and dust component(s), we performed SED modeling. This is challenging: the {\it HST}, NIRCam, and MIRI observations were obtained a few years apart (the most recent NIRCam data only $\sim$4 months before explosion), so describing the combined dataset with a single, time-independent model is not necessarily well justified, if the system is evolving and it is not known how such a system may appear immediately before explosion. We therefore applied only basic assumptions of a blackbody stellar component and simple, spherical dust-shell models (Sec.~\ref{sec:A_prog}). Under these assumptions, pure C-dust around a large evolved star provides a reasonable description of the combined SED; other dust compositions (silicates, metallic iron), while cannot be excluded, are much less probable based on the SED fitting (see details in Sec. \ref{sec:A_prog}).
At this distance, either an unresolved stellar blend or an underlying compact cluster may contribute to the observed radiation. However, we note the steep rise in the observed SED between 1.5 and 10~$\mu$m is not consistent with a $>$10-Myr old cluster (assumed to be a host of a Ia progenitor system), since they seem to be dust-poor with flat SEDs in this wavelength region (e.g., \cite{Whitmore_2025}, see their Fig. 7).
Late-time follow-up observations would be essential in providing stricter constraints.

\section{Discussion and conclusions}

As presented, i) the general shapes of the early-time spectra, ii) the strengths and line ratio of the Balmer lines, and iii) the early-phase LC evolution suggest that SN~2026sqf belongs to the rare type of SNe Ia-CSM. If so, this is by far the closest object from this class.

Moreover, SN~2026sqf could also be the first case of direct progenitor system identification of a Type Ia-CSM object: we clearly identified a source in pre-explosion {\it JWST}/NIRCam and MIRI images, and the source can also be (slightly) detected in {\it HST} F814W images. 
Given the astrometric uncertainty ($\lesssim$0.25\arcsec), we cannot exclude the possibility that a different, unrelated star within the alignment error circle is the true counterpart to SN~2026sqf. Among the sources present, however, we have identified the most promising candidate based on its photometric and physical properties.
The constructed SED of the potential progenitor system indicates a cool, large, luminous object surrounded by dense, dusty material, consistent with an extreme AGB star.

The presence of such a dense, H-rich CSM suggests substantial pre-explosion mass loss and makes binary interaction a plausible interpretation. Common-envelope (CE) evolution (CEE) phase -- when dynamically unstable mass transfer causes the primary star to engulf the secondary component in a shared common envelope \citep{Paczynski_1976,Webbink_1984} -- is a possible explanation, as it is also assumed for binaries with an AGB primary \citep{Sand_2020,Sarkar_2022}. 
A candidate for such a system was recently identified in M31 by \cite{Karambelkar_2025}; its luminosity, at least in the near-IR, closely matches that of the counterpart of SN~2026sqf (Fig. \ref{fig:sed_analogs}).
Although no observed SN Ia-like explosion has yet been directly connected to such a system, several scenarios link SNe Ia-CSM to a WD that explodes within a CE or merges with the core of its AGB companion \citep[e.g.,][]{Livio_2003,Soker_2013,Uno_2023b}. Our current observations, however, cannot distinguish the CEE scenario from other forms of enhanced or episodic mass loss in an interacting binary, such as winds driven by Roche-lobe overflow or recurrent nova ejections.

In summary, the data assembled here consistently point to an SN Ia-CSM arising from a WD–AGB binary, though late-time multiwavelength observations are needed to test and refine this picture. Continued JWST monitoring will be especially valuable: mid-infrared photometry and spectroscopy at later epochs can track the evolution of the dust temperature and mass as the SN shock sweeps up and heats the CSM shell, constraining the geometry and history of the pre-explosion mass loss. Combined with optical, radio, and X-ray follow-up and more detailed modeling, these observations offer a rare opportunity to fully characterize an SN Ia-CSM progenitor system. If the identification holds, SN 2026sqf will serve as a benchmark for the class as a whole.

\section*{Data availability}

Optical spectra are available at WISeREP\footnote{\href{https://www.wiserep.org/object/30976}{https://www.wiserep.org/object/30976}}.

\begin{acknowledgements} 
The authors thank the anonymous referee for their valuable comments.
This research is based in part on observations made with the NASA/ESA Hubble Space Telescope, obtained from the Space Telescope Science Institute (STScI), which is operated by the Association of Universities for Research in Astronomy (AURA), Inc., under NASA contract NAS 5–26555. 
This work is based in part on observations made with the NASA/ESA/CSA James Webb Space Telescope. The data were obtained from the Mikulski Archive for Space Telescopes (MAST) at STScI, which is operated by AURA, Inc., under NASA contract NAS 5-03127 for JWST. 
Observations from the NOT were obtained through the NUTS2 collaboration which is supported in part by the Instrument Centre for Danish Astrophysics (IDA), and the Finnish Centre for Astronomy with ESO (FINCA) via Research Council of Finland grant 306531. 
T.K. and U.P. acknowledge support from the Research Council of Finland project 360274.
T.-W.C. and A.A. acknowledge the financial support from the Yushan Fellow Program by the Ministry of Education, Taiwan (MOE-111-YSFMS-0008-001-P1) and the National Science and Technology Council, Taiwan (NSTC grant 114-2112-M-008-021-MY3). D.M. acknowledges support from NSF through grant AST-2206532. S.M. acknowledges financial support from the Research Council of Finland project 350458.
We thank J. Pierel for answering the questions that arose while using \texttt{space{\textunderscore}phot}.
\end{acknowledgements}

\bibliographystyle{aa}
\bibliography{Reference}

\begin{appendix}
\nolinenumbers

\section{Early-time spectra and photometry of SN~2026sqf}\label{sec:A_early}

\begin{figure}[!ht]
    \centering
    \includegraphics[width=0.7\columnwidth]{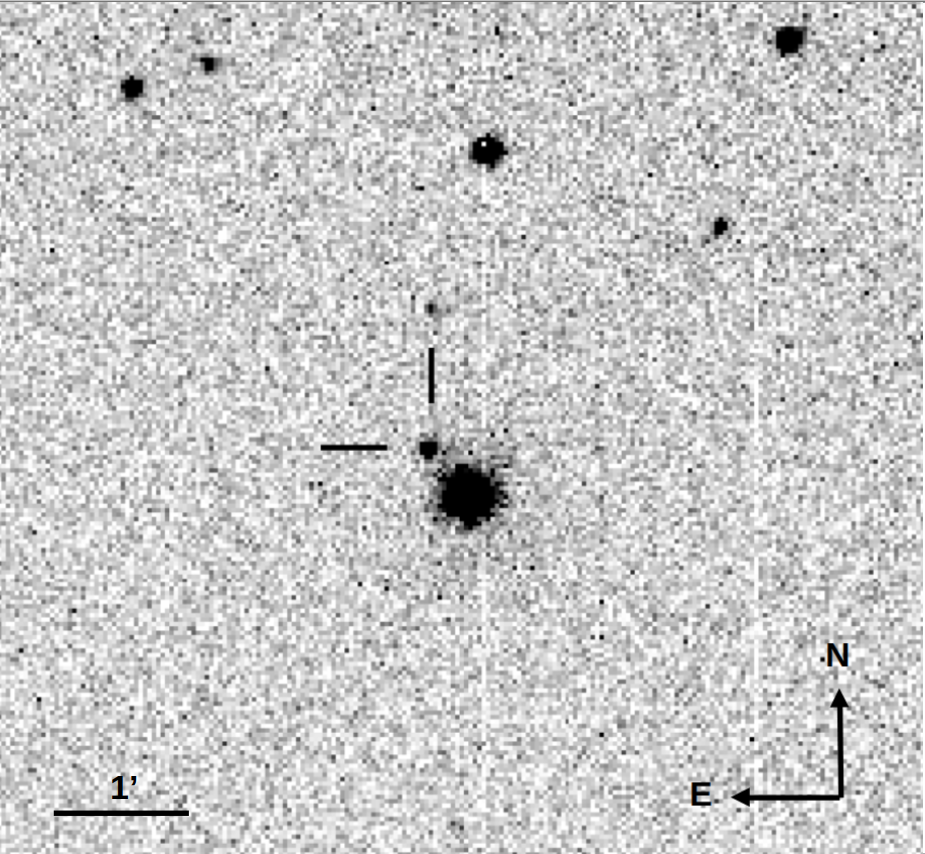}
     \caption{SN~2026sqf on P. Wiggins' discovery image obtained on 2026-07-08.25 UT near Erda, Utah, USA (0.35m Celestron C-14 telescope mounted with an SBIG ST-10XME camera, clear filter).}
    \label{fig:sn2026sqf}
\end{figure}

\subsection{Distance and reddening}\label{sec:A_early_dist}

Redshift-independent distances in NED show a large scatter between 10.8 and 18.7 Mpc and originate from decades-old studies applying the Tully-Fisher relation. Converting the redshift to luminosity distance (by assuming $H_0$ = 68.7 km s$^{-1}$ Mpc$^{-1}$, $\Omega_m$ = 0.31, and $\Omega_{\Lambda}$= 0.69, also considering the influence of the Virgo cluster, the Great Attractor, and the Shapley supercluster), results in 20.6 Mpc. Another estimation,
$17.8 \pm ^{0.6} _{0.4}$ Mpc, is based on the application of the Expanding Photosphere Method (EPM) on the Type IIP SN~2021gmj \citep{Meza-Retamal_2024} in the same host galaxy. Calculating an average of the two latter values, we use $D=19.0 \pm 1.5$ Mpc throughout our work for the distance.

Narrow \ion{Na}{i} D lines are not apparent in the early spectra either at $z$=0 or at the host's assumed redshift in our spectra (which could be due to the resolution limits of our data). Nevertheless, we assume here that the host reddening is not larger than the Galactic component \citep[$E(B-V)_\textrm{gal}=0.02$ mag][]{SF11} and apply an upper limit for the total extinction of $E(B-V)_\textrm{tot}=0.04$ mag.

\subsection{Spectroscopy}\label{sec:A_early_spec}

\begin{figure}[!ht]
    \centering
    \includegraphics[width=0.5\textwidth]{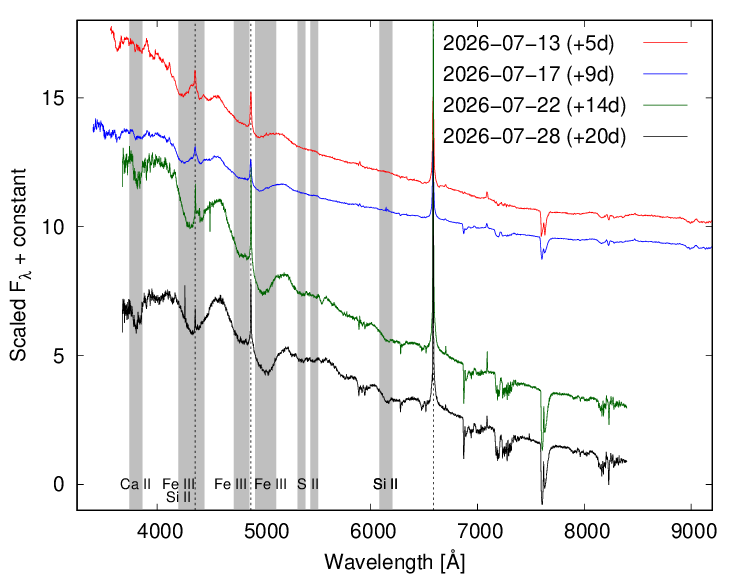}
    \includegraphics[width=0.45\textwidth]{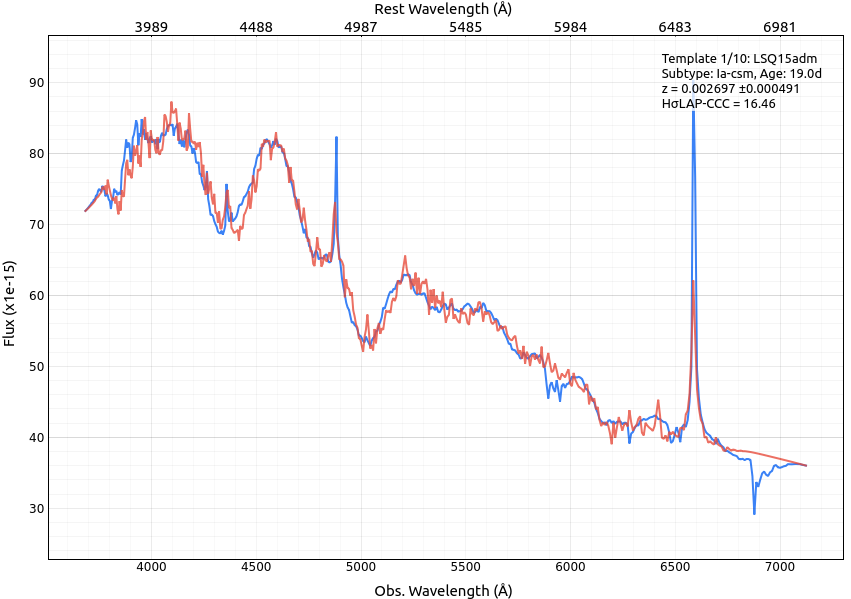}
     \caption{{\bf Top:} Early-time NOT/ALFOSC optical spectra of SN~2026sqf. Phases are given since the discovery date (2026-07-08.25 UT, 61229.25 MJD). Dashed vertical lines and grey stripes denote H Balmer emission lines and further prominent spectral features. The region between 6800$-$7800 \AA\ is dominated by telluric lines. {\bf Bottom:} 2026-07-28 (+20d) spectrum of SN~2026sqf compared to the best-match template, SN Ia-CSM LSQ15adm, in SNID-SAGE (with blue and red lines, respectively).}
    \label{fig:spec}
\end{figure}

A log of spectroscopic observations can be found in Table \ref{tab:spec}. 
The observed spectra were reduced -- bias- and flat-corrected, wavelength- and flux-calibrated -- using {\tt Foscgui} and {\tt IRAF} for grism 4 and 7/8, respectively. 
The results of our analysis of the early-time spectra are shown in Fig. \ref{fig:spec}.

When running \texttt{SNID-SAGE}, the observed spectra of SN~2026sqf match best with the Ia-CSM templates SNe~2005gj and LSQ15adm at all four epochs (resulting in H$\sigma$ LAP-CCC $>$14 values, while best-match SN II templates reach only values of $\sim$ 9 and $\sim$ 6 before and after $\sim$ +10d, respectively); see bottom panel of Fig. \ref{fig:spec} for a comparison plot.

Before measuring EWs of H${\alpha}$ and H${\beta}$ lines, we fit the continuum level of spectra around the two lines using the \texttt{IRAF continuum} task (applying low-order polynomials). We then applied both direct integration and fit of single-component Voigt profiles to both H${\alpha}$ and H${\beta}$ lines with the \texttt{splot} task. Final EW values are from the means and standard deviations of the two different methods.

\begin{table}[!ht]
\begin{center}
\caption{Log of spectroscopic observations obtained with NOT/ALFOSC. Phases are given with respect to day of discovery.}
\label{tab:spec}
\begin{tabular}{cccccc}
\hline
\hline
Date & Phase & Grism & t$_\textrm{exp}$ & Range & R \\
(UT) & (days) & ~ & (s) & (\AA) & ($\lambda$/$\Delta\lambda$)\\
\hline
2026-07-13 & $+5$ & 4 & 600 & 3600$-$9200 & 360 \\
2026-07-17 & $+9$ & 4 & 450 & 3600$-$9200 & 360 \\
2026-07-22 & $+14$ & 7 & 300 & 3650$-$7110 & 650 \\
~ & ~ &  8 & 300 & 5680$-$8580 & 1000 \\
2026-07-28 & $+20$ & 7 & 300 & 3650$-$7110 & 650 \\
~ & ~ & 8 & 300 & 5680$-$8580 & 1000 \\
\hline
\end{tabular}
\end{center}
\smallskip
\end{table}

\subsection{Photometry}\label{sec:A_early_phot}

BVRI and g'r'i' photometry from the Supra Solem Observatory was obtained with a PlaneWave CDK 500 (0.5 m) telescope. The images were calibrated using the ACP Observatory Control software. 
We obtained three epochs of ugriz imaging with the SLT and LOT. Further details of the Lulin Observatory facilities, data reduction, and photometry pipelines are provided in \cite{Aryan_2025}.
Unfiltered observations from \cite{Wiggins_2026sqf_disc} were obtained with a 0.35-m Celestron C14 telescope. 

Template subtraction was performed on the g'r'i'z' science images using the Saccadic Fast Fourier Transform for image subtraction (SFFT, \citealp{Hu_2022}) with Pan-STARRS images used as templates. No template subtraction was applied to the u', BVRI, or unfiltered data. PSF photometry was performed on the subtracted images with \texttt{AutoPhOT} and calibrated from zeropoints derived from Gaia XP spectra \citep{Brennan2022, daniela2025}. For the unfiltered C14 images, the quantum efficiency curve of the SBIG ST-10XME camera (Kodak KAF3200E CCD) was used to approximate the effective passband for calibration.
A comparison of early-time LCs of SN~2026sqf to those of known SNe Ia-CSM can be seen in Fig. \ref{fig:LC_comp_ztf}.

\begin{figure}
    \centering
    \includegraphics[width=\columnwidth]{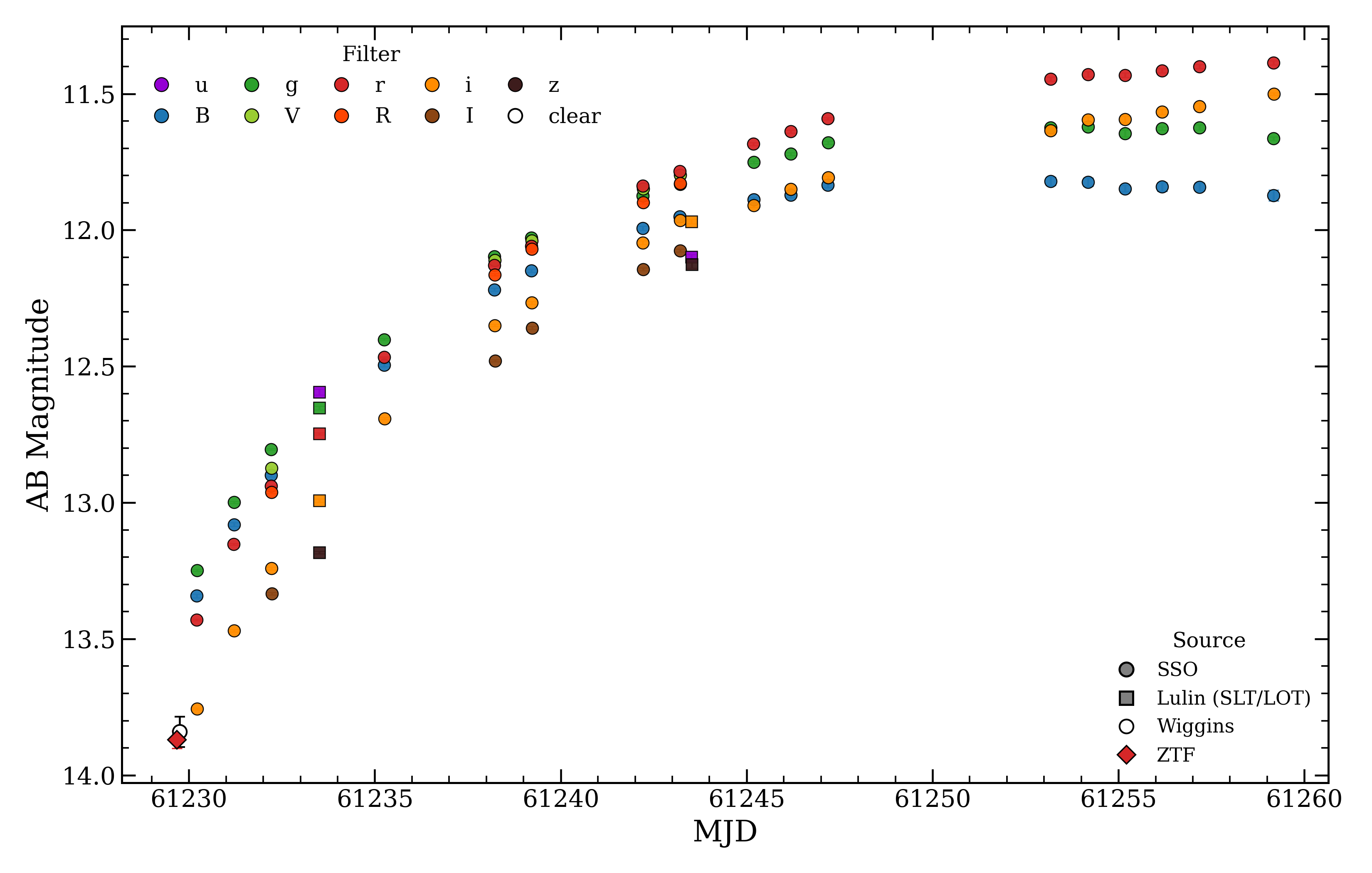}
     \caption{Photometry of SN~2026sqf. The SN was followed up to August 6, near maximum light, shortly before entering solar conjunction.}
    \label{fig:LC}
\end{figure}

\begin{figure*}[!ht]
    \centering
    \includegraphics[width=0.4\textwidth]{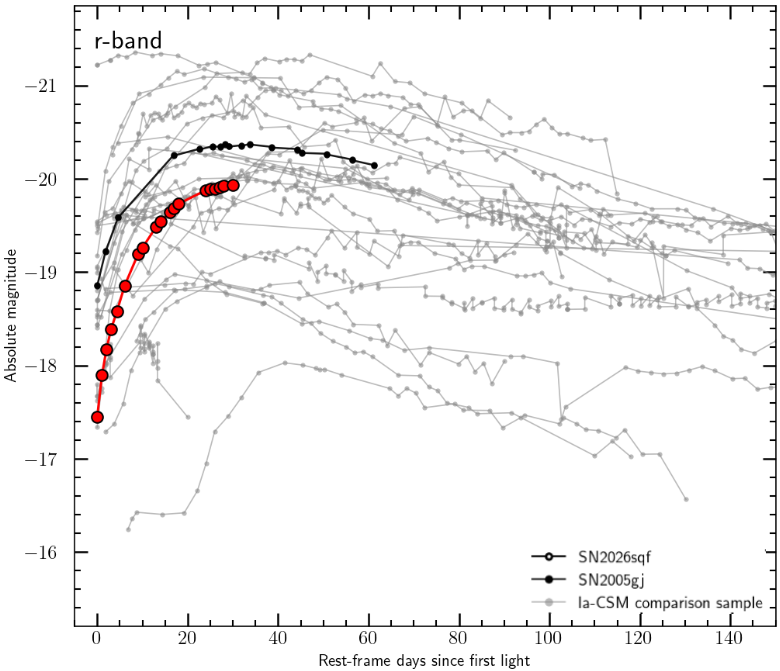}
    \includegraphics[width=0.55\textwidth]{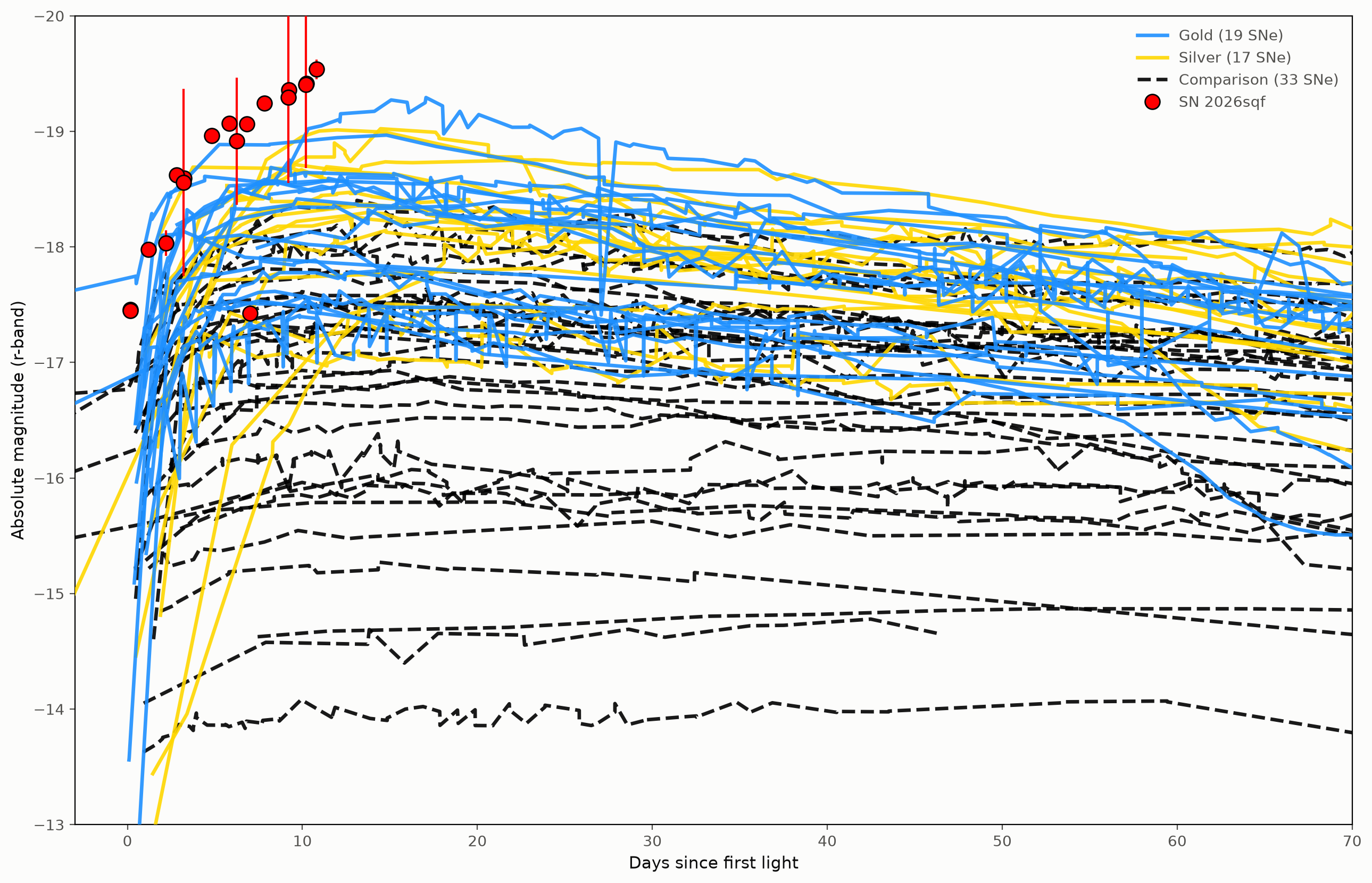}
     \caption{Early-time r-band LC of SN~2026sqf, compared to known SNe Ia-CSM from the ZTF sample \citep[left,][]{Terwel_2025a,Terwel_2025b} and to the sample of Type II SNe showing early-time interaction \citep[right,][]{Jacobson-Galan_2024}.}
    \label{fig:LC_comp_ztf}
\end{figure*}

\section{JWST and HST images and photometry of the potential progenitor system}\label{sec:A_jwst_hst}

{\it Image alignment and relative astrometry}: 
We used the JWST HST Alignment Tool \citep[JHAT;][]{Rest_2023} to initially align {\it JWST}, {\it Spitzer}, and {\it HST} images with each other. 
We used IRAF/PyRAF {\tt geomap} to establish the astrometric solutions for 8 stars on stacked SSO R-band images and compared it to the pre-explosion {\it HST} ACS F814W and WFC3/UVIS F814W images, independently. 
For ACS F814W, the $1\sigma$-uncertainty that we obtained for all fiducial stars was 2.09 (WFC) pixels (0.105\arcsec); removing one star each of 8 trials results in a range of 1.58 to 2.25 pixels (0.079--0.113\arcsec) with an rms of 0.26 pixels (0.013\arcsec). For WFC3/UVIS F814W, the numbers are 3.89 UVIS pixels (0.154\arcsec), and a range of 2.81 to 4.19 pixels (0.111--0.166\arcsec) with an rms=0.54 pixels (0.021\arcsec), respectively. Within these error circles, we identified a candidate progenitor in F814W. Bootstrapping the positions both of the SN and the candidate from the ACS image, we obtained a positional uncertainty of 0.47 pixels (0.015\arcsec), and a range from 0.38 to 0.49 pixels (0.011--0.015\arcsec) with an rms=0.04 pixels (0.001\arcsec), respectively. To within the uncertainties, the candidate in F814W has a direct counterpart in the NIRCam image.

{\it JWST/NIRCam and MIRI photometry}: 
In \texttt{space{\textunderscore}phot}, we performed point-spread-function (PSF) photometry on background-subtracted level-two data products using the implemented functions of \texttt{WebbPSF} \citep{Perrin_2014}. In order to calibrate the flux, we applied offsets by measuring the PSF of all the stars in the field and comparing them to the corresponding catalogs created by the pipeline. 
The fluxes of all four dithers of each filter were then averaged. 
In MIRI images, the complex sky background may significantly affect the reliability of the photometry; thus, for estimating the background levels, we used selected sky regions and subtracted their average values.
We repeated both NIRCam and MIRI photometry with {\tt Dolphot} \citep{Dolphin2016} implemented for {\sl JWST}, following the parameters settings and methods recommended by \citet{Weisz2024}; see a more detailed description of the method, e.g., in \cite{SVD_2026_25pht}. 
The final AB magnitudes and converted fluxes (as described in Sec. \ref{sec:data_jwst_hst_phot}) are shown in Table \ref{tab:JWST_phot}.

\begin{figure}[!ht]
    \centering
    \includegraphics[width=0.5\textwidth]{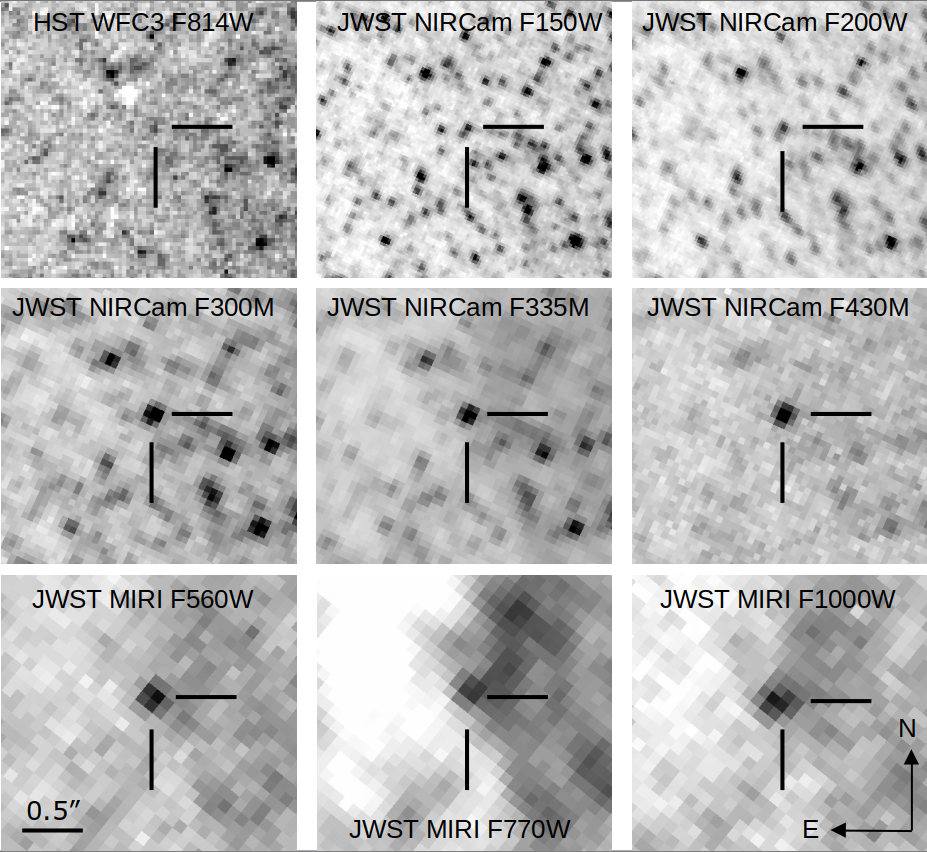}
     \caption{The potential progenitor system of SN~2026sqf on HST WFC3/UVIS (2021-12-26), JWST/NIRCam (2026-03-20) and MIRI (2024-04-12) images. North is up, east is to the left.}
    \label{fig:miri}
\end{figure}

Note that for the shortest wavelengths (F150W and F200W), in a southeast direction, another source is seen close to the suspected progenitor star (see Fig. \ref{fig:miri}). We found this other source to be $\sim$0.3 and $\sim$0.9 mag fainter in F150W and in F200W than the candidate, respectively, and not detectable in F300M and longer-wavelength NIRCam bands; thus, it probably does not significantly affect the photometry from the MIRI data (where the spatial resolution would be not high enough to separate the two objects).

{\it HST photometry}: 
The {\tt Dolphot} photometric analysis of the available {\sl HST\/} data followed the methods described by \citet{VanDyk2024}; in short, we pre-processed both the individual WFPC2 frames with {\tt Astrodrizzle} \citep[][with the additional benefit of flagging cosmic-ray hits in the data]{STScI2012b} and then measured the photometry. The source measured at the candidate position for both ACS and WFC3 is considered stellar (object\_type=1) by Dolphot.
The detection upper limits for the HST bands other than F814W are at $5\sigma$ and were estimated in the same way and following the same logic as described by \citet{VanDyk2023}.
During \texttt{space{\textunderscore}phot} photometry of WFC3 and ACS images, we followed the description found on the program's website\footnote{\href{https://space-phot.readthedocs.io/en/latest/api.html}{https://space-phot.readthedocs.io/en/latest/api.html}}. We carried out aperture photometry on Level-2 flc images (calibrated and flat-fielded single exposures corrected for charge transfer efficiency (CTE)); we used 2, 3, and 5 pixel values for the aperture radius, and the inner and outer radii of the annulus, respectively.
Regarding {\it HST} images, we have a slight ($\sim$ 4 $\sigma$) detection of a source found at the same position as on {\it JWST}/NIRCam and MIRI images. We only have upper limits for all the other filters (see Table \ref{tab:HST_phot}).

{\it Chance-projection test:} Following the referee's suggestion, we also performed a test of chance-projection of the SN position with a similar source across the face of the host galaxy. We considered that extent as the total isophotal radii and position angle of the host from \cite{Jarrett_2000} via NED. Following a method similar to that described by \cite{TWChen_2026}, we estimated the surface densities of all stellar sources (Dolphot object\_type parameter = 1) across the host in  ACS F814W, WFC3 F814W, and NIRCam F150W images that were $\pm1\sigma$ of the measured brightness for the candidate.
The areas of the 3$\sigma$ error circles for ACS, WFC3, and NIRCam F150W are then 0.31, 0.67, and 0.006\arcsec$^2$, respectively, and we computed chance-projection probabilities of 0.056, 0.134, and 6.6$\times 10^{-4}$ for any source other than the candidate being detected at the SN position. Moreover, we emphasize that no other stellar source in the host galaxy has the same SED, to within 3$\sigma$ and across all of NIRCam, as the candidate.

\begin{figure*}
    \centering
    \includegraphics[width=0.9\textwidth]{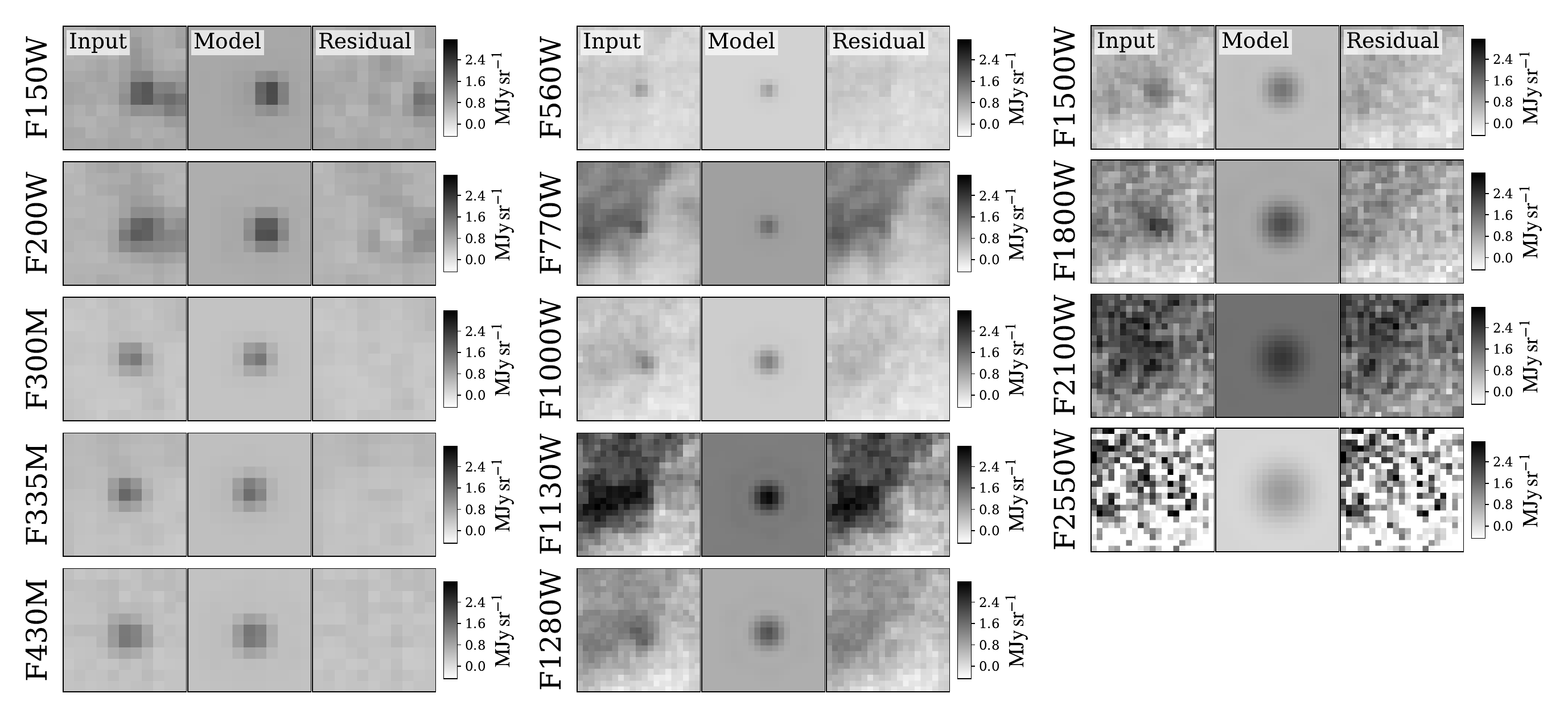}
     \caption{Output of \texttt{space{\textunderscore}phot} for pre-explosion {\it JWST} photometry of the potential progenitor system: observed data, PSF models, and residual images for all the filters. Orientation is different than in Fig. \ref{fig:miri} since \texttt{space{\textunderscore}phot} fits in the XY plane of the drizzled images.}
    \label{fig:spacephot}
\end{figure*}

\begin{table}
\centering
\caption{{\it JWST}/NIRCam and MIRI photometry of the potential progenitor system of SN~2026sqf \label{tab:JWST_phot}}
 \renewcommand{\arraystretch}{1.5}
\begin{tabular}{lll} 
\hline
\hline
Filter & Magnitude & $F_{\nu}$ \\
~ & AB mag & ($\mu$Jy) \\
\hline
\multicolumn{3}{c}{NIRCam (GO 9246, 2026-03-20, PI: A. Lu)} \\
F150W & 25.31 $\pm$ 0.13 & 0.27 $\pm$ 0.03 \\
F187N & $>24.40$ & $<0.63$ \\
F200W & 24.75 $\pm$ 0.10 & 0.46 $\pm$ 0.04 \\
F300M & 24.09 $\pm$ 0.09 & 0.83 $\pm$ 0.07 \\
F335M & 23.84 $\pm$ 0.10 & 1.05 $\pm$ 0.10 \\
F430M & 23.60 $\pm$ 0.13 & 1.32 $\pm$ 0.16 \\
\multicolumn{3}{c}{MIRI (GO 3295, 2024-04-12, PI: D. Sand)$^a$} \\
F560W & 23.10 $\pm$ 0.15 & 2.09 $\pm$ 0.29 \\
F770W & 22.40 $\pm$ 0.30 & 3.98 $\pm$ 1.10 \\
F1000W & 22.10 $\pm$ 0.15 & 5.25 $\pm$ 0.73 \\
F1130W & $>21.30$ & $<10.96$ \\
F1280W & $>21.57$ & $<8.55$ \\
F1500W & $>21.35$ & $<10.47$ \\
F1800W & $>20.72$ & $<18.71$ \\
F2100W & $>20.95$ & $<15.13$ \\
F2550W & $>20.43$ & $<24.43$ \\
\hline
\end{tabular}
\tablefoot{$^a$ Publicly available data, downloaded from MAST.}
\end{table}

\begin{table}
\caption{{\it HST} WFPC2, ACS, and WFC3 photometry of the potential progenitor system of SN~2026sqf \label{tab:HST_phot}}
 \renewcommand{\arraystretch}{1.5}
\begin{tabular}{llll} 
\hline
\hline
Filter & \multicolumn{2}{c}{Magnitude} & $F_{\nu}$ \\
~ & Vega mag$^a$ & AB mag & ($\mu$Jy) \\
\hline
\multicolumn{4}{c}{WFPC2 (GO 6639, 1997-03-13, PI: G. Meurer)} \\
F336W & $>$23.4 & $>$24.6 & $<$0.52 \\
F439W & $>$24.4 & $>$24.2 & $<$0.76 \\
F814W & $>$24.3 & $>$24.7 & $<$0.48 \\
\multicolumn{4}{c}{WFPC2 (GO 8645, 2000-09-12, PI: R. Windhorst)} \\
F300W & $>$24.1 & $>$25.4 & $<$0.25 \\
F814W & $>$24.3 & $>$24.7 & $<$0.48 \\
\multicolumn{4}{c}{ACS/WFC, (GO 9892, 2003-10-21, PI: R. Jansen)} \\
F625W & $>25.0$ & NaN$^b$ & NaN$^b$ \\
\multicolumn{4}{c}{WFC3/UVIS (GO 16691, 2021-12-26, PI: R. Foley)} \\
F625W & 26.36 $\pm$ 0.22 & $>26.27$ & $<0.11$ \\
F814W & 25.81 $\pm$ 0.22 & 26.25 $\pm$ 0.22 & 0.115 $\pm$ 0.020 \\
\multicolumn{4}{c}{ACS/WFC (GO 17070, 2023-10-08, PI: C. Kilpatrick)} \\
F555W & $>26.3$ & $>28.05$ & $<0.02$ \\
F814W & 26.06 $\pm$ 0.20 & 26.50 $\pm$ 0.20 & 0.091 $\pm$ 0.015 \\
\multicolumn{4}{c}{WFC3/UVIS (GO 17506, 2024-09-30, PI: W. Jacobson-Galan)} \\
F275W & $>24.8$ & NaN$^b$ & NaN$^b$ \\
F555W & 26.61 $\pm$ 0.15 & $>26.34$ & $<0.11$ \\
\hline
\end{tabular}
\tablefoot{All data were downloaded from MAST. NGC~3310 was also captured during WFPC2 GO 5479 and WFC3/UVIS GO 14762 programs, but the SN site is outside the field of views. $^a$ For ACS and WFC3 data, we used \texttt{space{\textunderscore}phot}, which convert fluxes directly to AB magnitudes. $^b$ Negative fluxes.}
\end{table}

\section{Modeling the SED of the potential progenitor system}\label{sec:A_prog}

In our spherical circumstellar dust-shell model, the total flux from dust of mass $M_{\rm d}$ emitting thermally at a single equilibrium temperature $T_{\rm d}$, located at a distance $d$, is
\begin{equation}
F_{\nu} = \frac{M_{\rm d}B_{\nu}(T_{\rm d})\kappa_{\nu}P_\textrm{esc}(\tau)}{d^2}, 
\end{equation}

\noindent where $B_{\nu}(T_{\rm d})$ is the Planck function and $\kappa_{\nu}$ for a range of published optical constants for carbonaceous \citep{Jager_1998} and silicate species \citep{Draine_2007}, using both Mie theory and, alternatively, a Rayleigh-limit continuous distribution of ellipsoids (CDE2; Eq. 28 of \citet{Draine_2021}). 
$P_\textrm{esc}(\tau)$ is the escape probability of the infrared photons from the emitting region and can be given for a homogeneous dusty sphere as:

\begin{equation}
P_{esc}(\lambda) = \frac{3}{4\tau_{\lambda}} \Big[1 - \frac{1}{2\tau_{\lambda}^2} + \left (\frac{1}{\tau_{\lambda}} + \frac{1}{2\tau_{\lambda}^2} \right)e^{-2\tau_{\lambda}} \Big]\, ;
\end{equation}

\noindent in the optically thin cases ($\tau <<$1), $P_\textrm{esc}(\tau) \approx 1$. We used a grain radius of $a = 0.15~\mu\text{m}$ \citep[after][]{Groenewegen_2018}.

Extinction was also applied as two independent screens following the \citep{Gordon_2023} law with $R_V = 3.1$, adopting $E(B-V)_{\rm MW} = E(B-V)_{\rm host} = 0.02~\text{mag}$ ($A_V^{\rm tot} \simeq 0.125~\text{mag}$). Model fluxes were convolved with the SVO filter transmission curves (using the \texttt{astroquery} package), and non-detections were treated as one-sided constraints on the residuals.

\begin{figure}
    \centering
    \includegraphics[width=0.85\columnwidth]{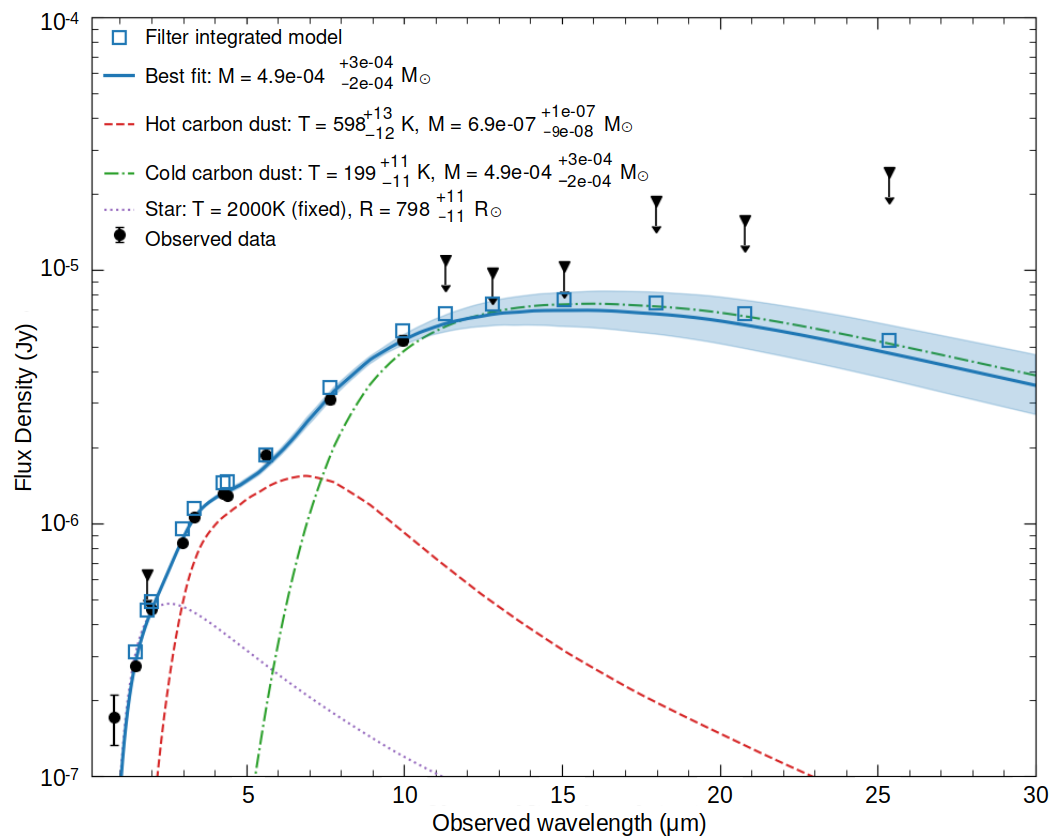}
    \includegraphics[width=0.85\columnwidth]{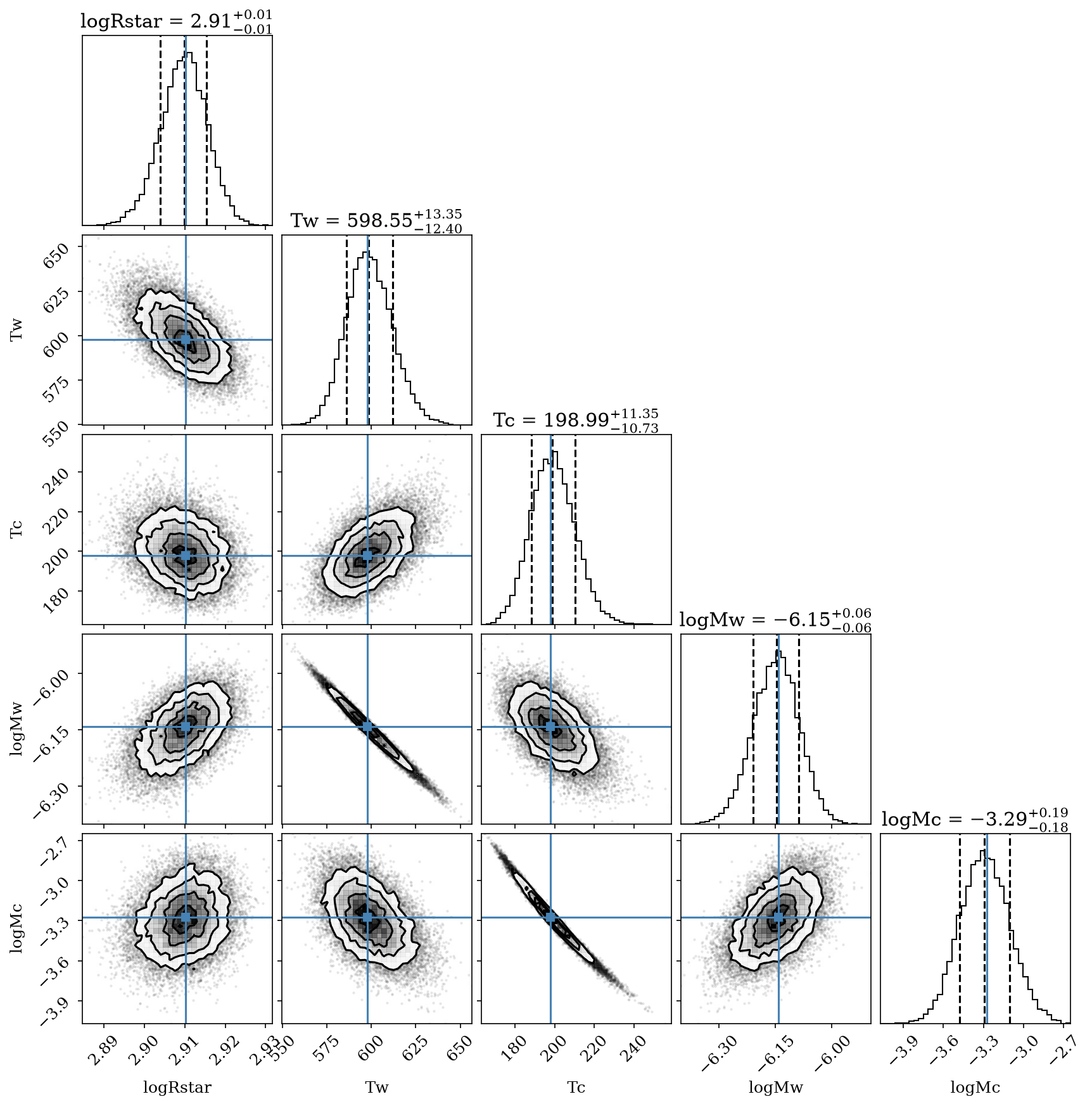}
    \includegraphics[width=0.85\columnwidth]{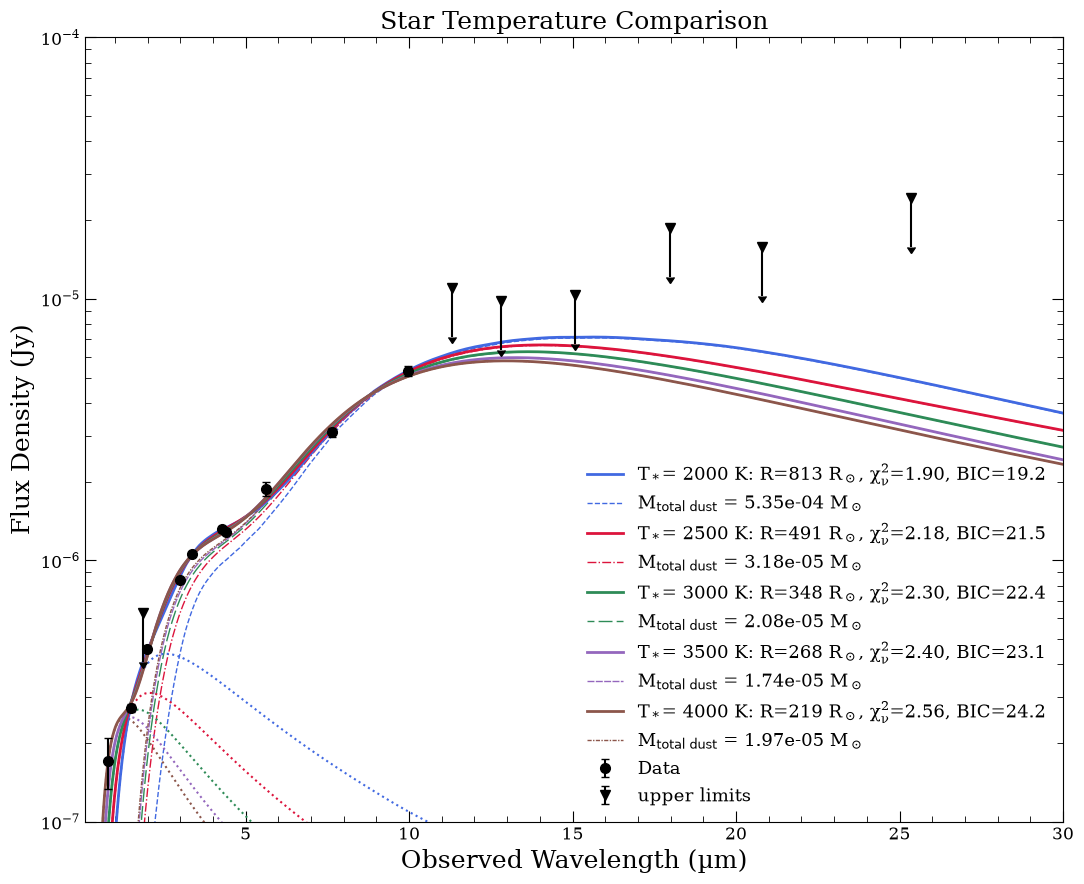}
    \caption{{\bf Top:} One of our three-component (star + warm dust + cold dust) models of the {\it HST} + {\it JWST}/NIRCam and MIRI SED of the potential progenitor system of SN~2026sqf. {\bf Middle:} Corner plot of the model fitting above. {\bf Bottom:} Model fits using various combinations of stellar temperatures (T$_{*}$) and radii (R$_{*}$).}
    \label{fig:SED_model}
\end{figure}

The full SED ({\it HST} + {\it JWST}/NIRCam and MIRI) comprises 16 photometric points: 9 detections from F814W to F1000W, and 7 upper limits (F187N and MIRI F1130W--F2550W). The flux density rises by a factor of ~20 in the SED from 1.5~$\mu$m to 10~$\mu$m, which cannot come from an extinguished stellar blackbody with $T_{*} \geq 2000$~K alone and requires an infrared excess from dust. Conversely, a dust-only model with $T_{\rm w}\leq800$~K fits substantially worse than one including a $\sim2000$~K stellar component ($\Delta\text{BIC} \approx43$), which we attribute to a stellar photosphere. Therefore, we modeled the SED as an extinguished stellar blackbody plus warm and cool optically thin dust components. Fixing $T_{*}$ between $2000$~K and $4000$~K and the host extinction, we allowed the stellar radius $R_{*}$ to vary between $0.1$ and $31{,}600~R_{\odot}$ while allowing the warm and cool dust temperatures $T_{\rm w}$ and $T_{\rm c}$, and the dust masses $M_{\rm w}$ and $M_{\rm c}$ to vary as well.
Best-fit parameters were obtained for each dust species by Levenberg-Marquardt minimization of the weighted residuals using the python package \texttt{lmfit}, and posterior distributions were sampled with the python package \texttt{emcee} where quoted uncertainties correspond to the 16th and 84th percentiles of the marginalized posteriors.
A set of our models and corner plots, are shown in Fig. \ref{fig:SED_model}.

Among the species considered, amorphous carbon (`cel400`; \citealp{Jager_1998}) gives the best fit ($\chi^2 = 22.9$ for both Mie and CDE2 opacities), and other amorphous carbons ('cel600', 'cel800', and 'cel1000') give $\chi^2=26.2$--$26.8$. Silicate-bearing models fit considerably worse, with the best of them, astronomical silicate \citep{Draine_2007} with CDE2 opacities, gives $\chi^2=332$. The other species, amorphous pyroxene (Mg$_{0.7}$Fe$_{0.3}$SiO$_3$; \citealp{Dorschner_1995}), silica (SiO$_2$; \citealp{Kitamura_2007}), and corundum (Al$_2$O$_3$; \citealp{Koike_1995}), all give $\chi^2 \gtrsim1600$. All models have the same number of free parameters, so the difference of $\Delta\chi^2\approx300$ means that silicates are strongly disfavored within our model. These models perform poorly even when $T_{*}$, $A_V^{\rm host}$ and $T_{\rm w}$ are left free (up to $T_{\rm w}=2000$~K). While the best-fitting species across $T_{*}=2000$--$4000$~K is always amorphous carbon ($\chi^2=22.9$--$30.7$), metallic iron grains \citep{Henning_1996}, which also have a featureless mid-infrared opacity, fit almost as well ($\chi^2=25.9$). We interpret this result as the photometry's tendency to favor a featureless dust continuum, and prefer carbon over iron, since the required presence of iron grains to reproduce this SED are not as physically feasible, as they require significantly more mass to produce the same SED as amorphous carbon grains.

The comparison also depends on the optically thin assumption. At the blackbody equilibrium radii, the best-fit dust masses imply optical depths $\tau_V$ of $\approx56$ (warm) and $\approx490$ (cool), and $\tau(10~\mu{\rm m}) \approx 0.6$ and $\approx5$. At these optical depths, the 9.7~$\mu$m silicate feature can be self-absorbed, so broadband photometry alone cannot exclude an optically thick silicate shell. We therefore describe the source as a dusty evolved-star (AGB) candidate. Carbon-rich dust is one plausible interpretation of its featureless mid-infrared continuum, but not a firm identification, where mid-infrared spectroscopy and follow-up modeling that include more complex radiative-transfer scenarios are needed to more confidently establish the composition. 

To assess the effect of the non-simultaneous photometry, we also fitted only the HST F814W and NIRCam photometry (six detections and the F187N upper limit), excluding MIRI, with a photosphere and a single dust component (three free parameters; $\chi^2 = 19.4$, dominated by F814W). For the same fixed photosphere ($T_* = 2000$~K, $A_V^{\rm host} = 0.062$), the results are indistinguishable from the full-SED fit: $R_* = 812~R_{\odot}$ and $L_* = 9.5\times10^{3}~L_{\odot}$ in both cases, and $T_w = 600$~K and $M_w = 7.0\times10^{-7}~M_{\odot}$ versus $598$~K and $7.2\times10^{-7}~M_{\odot}$. Treating the MIRI fluxes as upper limits in this fit leaves the solution unchanged, because the NIRCam-only model lies below all MIRI measurements. The stellar luminosity and radius and the warm dust are therefore determined by the $0.8$--$4.4~\mu{\rm m}$ data alone and are independent of the MIRI epoch. 

Extrapolated to the MIRI bands, this model underpredicts the observed fluxes by factors of 1.3, 2.1, and 5.5 at F560W, F770W, and F1000W respectively, with the discrepancy growing monotonically with wavelength. Attributing this excess to variability would require a factor of $\sim5$ brightening at $10~\mu{\rm m}$, but only $\sim25\%$ at 5.6~$\mu{\rm m}$, between epochs. Such a wavelength-dependent change is well beyond the $\lesssim1$ mag mid-infrared amplitudes of pulsating AGB stars \citep[e.g.,][]{LeBertre1992,LeBertre1993,Whitelock2006,Nowotny2011} and inconsistent with a simple luminosity change, which would scale all bands together. The cool component is thus required by the MIRI detections, although its mass could be biased at the factor-of-two level by inter-epoch variability. Without the MIRI detections, amorphous carbon, metallic iron, and astronomical silicate opacities fit the near-infrared data comparably ($\chi^2=19.4$, $20.5$, and $21.8$ respectively), whereas Mg-Fe pyroxene is already disfavored ($\chi^2=52.6$). The compositional constraint on the remaining species therefore rests on F770W and F1000W. In particular, the astronomical-silicate model fitted to the near-infrared excess predicts a strong 9.7~$\mu{\rm m}$ emission feature in the optically thin limit, overpredicting F1000W by a factor of 2.5 while matching F770W to within 13\%, whereas the observed rise from F770W to F1000W is smooth. 
We finally note that $R_*$ depends mainly on the single F814W measurement, which the model overpredicts by $\sim2.5\sigma$; therefore the derived stellar parameters should be regarded only as indicative within the adopted model.

\end{appendix}

\end{document}